\documentclass[prl,showpacs]{revtex4}

\usepackage{graphicx}
\usepackage{rotating}
\usepackage{amsmath}
\usepackage{amsfonts}
\usepackage{amssymb}
\usepackage{enumerate}
\usepackage{longtable}
\usepackage{dcolumn}% Align table columns on decimal point
\usepackage{bm}

\newcommand{\be}{\begin{equation}}
\newcommand{\ee}{\end{equation}}
\newcommand{\bn}{\begin{eqnarray}}
\newcommand{\en}{\end{eqnarray}}

\begin{document}

\title {Towards A Possible Charge-Kondo Effect in Optical Lattices}

\author {M. S. Laad$^{1}$,
L. Craco$^{2}$, and A. Taraphder$^{1,4}$}
\affiliation{ $^{1}$ Lehrstuhl f\"ur 
Theoretische
 Physik, Technische Universit\"at Dortmund, 44221 Dortmund, Germany\\
$^{2}$Max-Planck-Institut f\"ur Chemische Physik fester Stoffe, 01187 
Dresden,
Germany \\
$^{3}$Department of Physics and  Centre for
Theoretical Studies,\\ Indian Institute of Technology, Kharagpur 721302
India \\
$^{4}$Max-Planck-Institut f\"ur Physik komplexer Systeme, 01187 Dresden,
Germany
}
\date{\rm\today}

\begin{abstract}
The Kondo effect underpins a large body of recent developments in the
physics of $d$- and $f$-band compounds. Although its  {\it charge} analog
is a rarity in solids, the recent observations of the charge Kondo effect
and the consequent rise in superconducting T$_c$ encourage a search for
other accessible systems. Motivated by the possibility of wilfully tuning
the {\it sign} of the inter-electronic interaction in optical lattices,
we study conditions for the elusive {\it charge Kondo liquid} (CKL) state
to manifest. We propose that a combination of Feshbach resonances and
sequentially controlled laser pulses may produce the CKL. We show that
the CKL is {\it never} a stable ground state, appearing only when the
ordered ground states are destabilized. Finally, we discuss interesting
analogies with nuclear matter.
\end{abstract}


\maketitle

The Kondo effect has historically played a monumental role in solid state
physics~\cite{ref1}, where quasi-bound state (singlet) formation involving
a local moment screened by the itinerant electron spins leads a 
renormalized
Fermi liquid (FL) with huge enhancements of quasiparticle masses, whence
the name ``heavy fermion''. Subsequently, more exotic, 
quadrupolar~\cite{ref2}
and orbital~\cite{ref3} versions were invoked in other contexts. The
destruction of FL behavior at quantum phase transitions near magnetic order
in some $f$-electron systems is thought to be linked to ``unbinding'' of
this Kondo quasi-bound singlet state~\cite{si}. However, the intriguing
possibility of observing a {\it charge} analogue of the Kondo effect,
requiring an {\it attractive}, local interaction, has not been addressed
carefully. In practice, materials such as $Ba_{1-x}(K,Pb)_{x}BiO_{3}$ are
modelled by variants of $U<0$ Hubbard model~\cite{varma,AT_95,AT_96,wilson}.

Recent experimental realization~\cite{gebal} of the {\it charge}
Kondo effect (CKE), predicted theoretically some time back~\cite{AT_91},
and the consequent rise of superconducting $T_{c}$ in doped
$PbTe$~\cite{dzero} and $SnTe$ \cite{gebal}, has re-ignited interest in the
$U<0$ models. Here, chemical doping ($\le 2\%$) generates a narrow impurity
band in the semiconductor band gap, facilitating use of a $U<0$ lattice
models in the intermediate-to-strong coupling limit ($|U|/W_{imp}>1$). The
interesting issues for theory are thus:
{\it What are the precise conditions under which a charge-Kondo Fermi liquid
state (the CKL state) can appear? Is it ever a stable phase at zero 
temperature?
Might a quantum phase transition (QPT), reminiscent of its well-studied spin
analogue, occur, and if so, how?  What instabilities might one generically
expect to arise near such a QPT?}

Advances in artificially engineered fermionic/bosonic optical lattices
allowing wilful manipulation of parameters open up yet another route to test
model Hamiltonian predictions.  Such systems may hold more promise for CKE,
given the paucity of real materials exhibiting $U<0$. In contrast to real
materials, these systems are free of inhomogeneities. In this context,
various interacting models have already been
realized~\cite{georges, esslinger}. The physical conditions for this lie
well within the regime of experimental techniques available. In practice,
different hyperfine states of same or different atomic species (acting,
as it  may, as different fermionic/bosonic species) can be trapped and
controlled independently. A particularly attractive feature is their
ready tunability: it is even possible to choose the sign of the interaction
in such systems; a mixture of two fermionic species interacting via an
attractive interaction is achieved by forcing a mixture of two such atomic
spin states through a Feshbach resonance, whence a bound state appears in
the two particle problem. A p-wave Feshbach resonance~\cite{regal} can
even create a {\it tunable} asymmetry in the interactions, allowing one
to access more complex ``hidden'' ordered states. Further, the periodic
potential of each species of atoms was independently tuned in an optical
lattice~\cite{mandel}. This could facilitate observation of the band
insulator (driven by staggered periodic lattice potential) to Mott
insulator (due to on-site interaction, $U$) transition, and possible
emergence of correlated metallic phases sandwiched between them.

Optical lattices thus provide a unique tool to address interesting questions 
posed above. A suitable model for a two-component fermionic system (having 
an attractive two-body interaction), {\it with} temporally separated laser 
pulses simulating inter-site {\it hybridization} between the two spinless
fermionic species ($a,\, b$) is described by the Hamiltonian

\bn
H = \sum_{<i,j>}t_{ab}(a_{i}^{\dag}b_{j}+h.c.) - U_{ab}\sum_{i}n_{ia}n_{ib}
+ \Delta_{0} \sum_{i}(n_{ia}-n_{ib})+\sum_{i,\alpha=a,b}
\epsilon_{\alpha} n_{i\alpha
\label{eq1}
}
\en

The hoppings are $t_{aa}\, (t_{bb})$ for $a-a\, (b-b)$ hopping and $t_{ab}$
for $a-b$ hopping between nearest neighbors. $\Delta$ is the
``charge transfer'' term related to on-site fermionic energies. In typical
optical lattice systems, the model parameters expressed in terms of the
recoil energy are, hopping 
$t=\frac{4E_{R}}{\sqrt\pi}({\frac{V_0}{E_R}})^{3/4}\,
e^{-2\sqrt{(V_{0}/E_{R})}}$ and local correlation $U\simeq 
E_{R}\sqrt{8/\pi}\,\,
a_{s}k_{L} (V_{0}/E_{R})^{3/4}$~\cite{georges}. Here, the recoil energy,
$E_{R}=\hbar^{2}k_{L}^{2}/2m$ is typically of order a few micro Kelvin.
($k_{L}$ and $V_{0}$ are the wave-vector and intensity of the laser
respectively). $a_{s}$ is the s-wave scattering length, and the fermionic
bandwidth, $W<<E_{R}$.

Typically, $t/\hbar \simeq 1$ kHz and $U/\hbar \simeq 20-30$ kHz
\cite{esslinger}. So $U/W\simeq O(2-3)$ in three dimensions. We envisage
comparable values for the hoppings and $U_{ab}$ in our case. This is
precisely the intermediate to strong coupling regime of the $U_{ab}<0$
``Hubbard''-like model that we study below. Focussing solely on possible
Kondo-like physics, we use dynamical mean-field theory (DMFT) to study
the model. Its reliability for understanding quasi-local collective Kondo
coherence in correlated metals is already established~\cite{kotliar-06}. 
Here, we show how a combination of experimental manipulations in optical
lattices facilitates observation of the elusive CKL state.

Eq.~(\ref{eq1}) is a generalized {\it negative}-$U$ Hubbard model
(relabeling  $a\rightarrow c_{\uparrow}, b{\rightarrow}c_{\downarrow}$)
with ``spin''-dependent hoppings, and a ``magnetic field''.  Given the
isomorphism between the $U>0$ Hubbard model and the $U<0$ Hubbard
model in a magnetic field (seen by performing a particle-hole
transformation), the possibility of the {\it charge} Kondo effect (CKE)
involving pair-pseudospin-quenching by itinerance manifests itself (the
pseudospin $\tau_{i}^{\alpha}=c_{ia}^{\dag}\sigma^{\alpha}_{ab}c_{ib}$).
On the other hand, lack of pseudospin SU$(2)$ invariance of $H$ (lowered
to Z$_{2}\times$U$(1)$) generically favors {\it ordered} ground states,
either a charge-density wave (CDW) or a superfluid (SF). A CKL state can
then appear as a {\it ground} state only upon quantum melting of these
ordered states, or for model parameters such that $T_{CKL}>T_{CDW,SF}$.

To proceed, we ``rotate'' the $a,b$ Fermions: writing
$f_{a}=(ua+vb)/\sqrt{u^{2}+v^{2}}, f_{b}=(va-ub)/\sqrt{u^{2}+v^{2}}$ with
$u=\sqrt{t_{aa}/(t_{aa}+t_{bb})}, v=\sqrt{t_{bb}/(t_{aa}+t_{bb})}$ on the
parameter curve $t_{ab}=\sqrt{t_{aa}t_{bb}}$ leads to

\bn
H  =  t\sum_{<i,j>}(f_{ia}^{\dag}f_{ja}+h.c)-
U_{ab}\sum_{i}n_{ifa}n_{ifb}
+  \Delta\sum_{i}(f_{ia}^{\dag}f_{ib}+h.c.) +
\sum_{i,\alpha=a,b}\epsilon_{\alpha}'n_{i\alpha}
\label{eq2}
\en
where $t=(t_{aa}+t_{bb}),\, \Delta=\Delta_{0}-uv$ and $\epsilon_{a}'=v^{2}
\epsilon_{a},\epsilon_{b}'=u^{2}\epsilon_{b}$. This is the $U<0$
Falicov-Kimball model (FKM) with a {\it local} hybridization~\cite{craco}.
If $\Delta$ is tuned to zero, this reduces to the pure FKM. Since the FKM
can be {\it exactly} solved in $D=\infty$~\cite{brandt}, we first solve
$H_{FKM}$ and then consider the effects of
$t_{ab}\ne \sqrt{t_{aa}t_{bb}}, \Delta\ne 0$.

We observe that $[n_{ifb},H_{FKM}]=0$ for {\it each} $i$, implying
a {\it local} U(1) invariance of $H_{FKM}$. This is {\it exactly}
the condition for degenerate, resonant charge fluctuations between local
configurations with $n_{f_{b}}=0,1$ to occur. Elitzur's theorem implies
that both local excitonic ($\langle f_{ia}^{\dag}f_{ib}\rangle$) and local
pairing ($\langle f_{ia}^{\dag}f_{ib}^{\dag}\rangle$) averages rigorously
vanish.Choosing the $f_{a}$-band density-of-states (DOS) to be a lorentzian
with half-width $W$; i.e, $\rho_{0}(E)=W/\pi(E^{2}+W^{2})$, the 
$f_{a}$-band
propagator within DMFT is {\it exactly} calculable as~\cite{si}
$G_{f_{a}}(\omega)=\frac{1-n_{fb}}{\omega+iW}+\frac{n_{fb}}{\omega+U_{ab}+iW}$,
where $n_{fb}=(1/N)\sum_{i}\langle f_{ib}^{\dag}f_{ib}\rangle$ with $N$
the number of lattice sites.  Correspondingly, the $f_{b}$-Fermion 
propagator
exhibits an infra-red singular, branch cut form, $G_{f_{b}}(\omega)\simeq
\theta(\omega)|\omega|^{-(1-\eta)}$ at low energy. Here,
$\eta$=tan$^{-1}(U_{ab}/W)/\pi$ is the so-called $s$-wave phase shift of
the $f_{a}$-band Fermions induced by the strong scattering term
($U_{ab}$) in $H_{FKM}$. The $f_{b}$-Fermion self-energy,
$\Sigma_{f_{b}}(\omega)\simeq |\omega|^{1-\eta}$, implying that the FL
quasiparticle residue $Z=0$, and the symmetry-unbroken metallic phase is
a non-FL metal. The underlying cause for non-FL behavior is explicit in
the impurity limit of $H_{FKM}$, where recoil-less (infinitely heavy)
$f_{b}$-fermions scatter the $f_{a}$-band fermions: it is the X-ray edge
problem, where such an infra-red singularity is known to occur. This
implies that the local {\it pairing} susceptibility (the analogue of the
``excitonic'' susceptibility for $U>0$) also diverges in the infra-red,
$\chi_{f_{a},f_{b}}'(\omega)=\int dt e^{i\omega t}\langle
f_{ia}^{\dag}f_{ib}^{\dag}(t);f_{ib}f_{ia}(0)\rangle \simeq
|\omega|^{-(2\eta-\eta^{2})}$: coupling of $f_{a}$-band fermions to this
singular pair susceptibility then destroys FL theory. Within DMFT, this
is an incoherent pair liquid state, where pair correlations diverge in
the infra-red {\it without} SC order. We dub this state the incoherent
Bose metal (IBM). The IBM is characterized by resonant charge fluctuations
as described above, and alluded to in earlier work~\cite{AT_91, das}. The
mixing of the degenerate single-particle and pair states reflect the
superposition of the ``emission" and ``absorption" initial states of the
(presently, attractive-U) X-ray edge problem. Since the $f_{b}$-fermions
are rigorously localized, the ``Fermi surface'' now wholly corresponds to
the dispersive $f_{a}$-band quasiparticles, with a smeared dispersion
reflecting unusually broad pseudoparticles, even at $T=0$. Depending on
$(t_{aa}+t_{bb})/U_{ab}$ and along the curve $t_{ab}=\sqrt{t_{aa}t_{bb}}$,
we generically find a {\it first order} transition near half-filling,
where the total fermion density, $n$, jumps from
$n_{<}=(n/2)+\frac{1}{\pi}tan^{-1}(U_{ab}/2W)$ to
$n_{>}=(3n/2)-\frac{1}{\pi}tan^{-1}(U_{ab}/2W)$.
For a bounded unperturbed DOS, numerical DMFT calculations
give results closely paralleling those described above~\cite{vandongen}.
At a critical coupling, $g<g_{c}=(U_{ab}/2W)_{c}$, this line of first-order
transitions ends in a second order {\it quantum critical} end point (QCEP)
at $T=0$~\cite{miyake}.

%At finite $T$, this unstable state is
%pre-empted, (i) either by CDW state below $T<T_{CDW}$
%with band-filling dependent periodicities (these break the translation
%invariance of $H_{FKM}$), signalled by divergent correlators
%$\chi_{ij}^{ab}=\langle n_{ifa}n_{jfb}\rangle$, or
%(ii) to a singlet superfluid away from half-filling.
%In the symmetry-{\it unbroken} phase, the IBM crosses over smoothly to a
%correlated Fermi liquid (see below), which turns out to be the
%pair-pseudospin-quenched analog of the well-known Kondo problem; i.e, a
%{\it charge} Kondo FL~\cite{AT_91}, upon small deviations from this
%critical parameter curve.

Away from $\Delta=0$ or $t_{ab}=\sqrt{t_{aa}t_{bb}}$, the terms,
$\Delta\sum_{i}(f_{ia}^{\dag}f_{ib}+h.c)$ (first case) or $
(t_{aa}t_{bb}-t_{ab}^{2})\sum_{<i,j>}(f_{ia}^{\dag}f_{jb}+h.c)$ (second
case) change the {\it local} U$(1)$ invariance of $H_{FKM}$ to a {\it
global} U$(1)$ invariance: only the {\it total} fermion number commutes
with $H$, $[n_{f_{a}}+n_{f_{b}},H]=0$.  This global symmetry can
indeed be (and is) spontaneously broken, and, away from ``half-filling'',
gives a {\it color} singlet superfluid (CSS) at low $T$, characterized by
$\Delta_{ab}=\sum_{i}\epsilon^{ab}\langle f_{ia}^{\dag}
f_{ib}^{\dag}\rangle>0$ below $T_{SC}=f(U_{ab}/t,\Delta,n)$. This is
readily checked by studying $H$ within the generalized Hartree-Fock random
phase approximation~\cite{AT_95}, which is valid now, since there is no 
local
constraint for any $\Delta\ne 0$. The enhancement of the superconducting
T$_c$ in doped $PbTe$ and $SnTe$ \cite{gebal} is now 
rationalized~\cite{dzero}
as a manifestation of the proximity to the resonant charge fluctuation 
regime
driven by strong quantum fluctuation between the two degenerate valence
states, $n_{f_{b}}=0,1$, envisaged in the original prediction~\cite{AT_91},
for $\Delta\simeq 0$ as above.
%The limit $|U_{ab}|<<t_{aa},t_{bb},t_{ab}$ is the usual BCS limit, while
%large $|U_{ab}|>>t_{aa},t_{bb},t_{ab}$
%leads to an anisotropic Heisenberg model, $H_{eff}=\sum_{<i,j>,
%\alpha,\beta}J_{\alpha\beta}S_{i}^{\alpha}S_{j}^{\beta}$ to second order in
%$t_{aa,bb,ab}/U_{ab}$.
%CDW order is then the ``magnetization'' ($\langle S_{i}^{z}\rangle$) 
%while the
%SS order, now interpreted as a Bose-Einstein condensate (BEC), 
%corresponds to
%XY order~\cite{varma,AT_95} in $H_{eff}$.
The respective transition temperatures can be computed from
DMFT~\cite{jarrell} over the full range of 
$|U_{ab}|/(t_{aa},t_{bb},t_{ab})$,
and yields a {\it smooth} BCS-BEC crossover with increasing $|U_{ab}|$.

This implies that the charge Kondo screening can {\it never} produce a
CKL {\it ground} state. However, {\it above} the ordering scales,
relevance/irrelevance of the $f_{a}-f_{b}$ hybridization determines whether
the IBM or a correlated FL metal will result. Consider the symmetry 
unbroken
phase for $\Delta>0$. The term $\Delta\sum_{i}(f_{ia}^{\dag}f_{ib}+h.c)$
hybridizes the ``infinitely heavy'' (for $\Delta=0$ above) $f_{b}$ fermion
with the $f_{a}$-band.  This is precisely the $U_{ab}<0$ periodic Anderson
model (PAM) with a {\it local} hybridization between the ``light'' ($a$)
and ``heavy'' ($b$) fermions.  To access the physics in this regime, we
solve this PAM within DMFT.

\begin{figure}[tbp!]
\begin{center}
\includegraphics[width=3.5in, height=2.5in, angle=0]{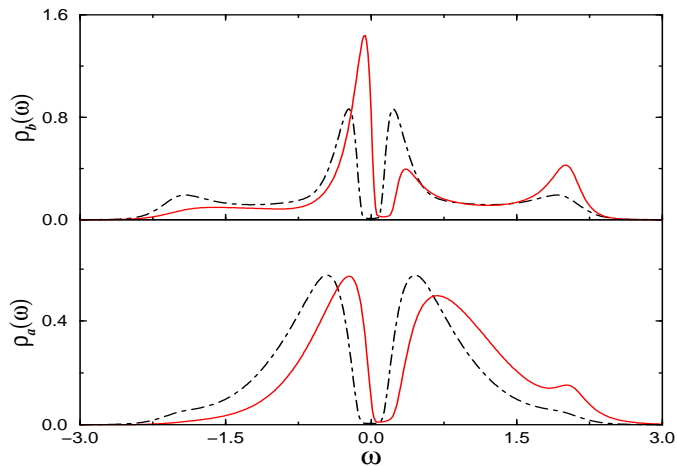}
\caption{%(Color online)
Local spectral function of the $U_{ab}<0$ periodic Anderson model within
DMFT in the symmetric (black, dashed) and asymmetric (red, dotted) limits
for the $a$ (lower panel) and $b$ Fermions (upper panel).
Upon varying $\epsilon_{a}'$, the charge Kondo insulator ($\epsilon_{a}'=0$)
melts into a charge-Kondo-FL state for $\epsilon_{a}'\ne 0$.}
\label{fig1}
\end{center}
\end{figure}

We choose a Gaussian unperturbed DOS for the $a$-fermions with 
$W/\hbar=12$kHz
and $|U_{ab}|/\hbar=20$kHz as representative values. To solve the
impurity problem of DMFT, we use the iterated perturbation theory (IPT) 
as an
impurity solver. Prior work carried out for the $U>0$ periodic Anderson 
model
with additional $f-d$ interaction~\cite{craco1} has demonstrated its
accuracy vis-a-vis the more ``exact'' QMC solver~\cite{jarrell}:  the IPT
spectral functions closely match those extracted from QMC at considerably
less numerical cost.  Additionally, self-energies can be readily extracted
from IPT in contrast to the difficulties encountered with QMC in this 
context.
 In Fig.~\ref{fig1}, we show the results obtained within DMFT for the PAM
for two cases: ($i$) $\epsilon_{a}'=0$, and ($ii$) $\epsilon_{a}' \ne 0$.
The first case has particle-hole symmetry, while the second does not.
For a half-filled $a$-band, in case (i), we find a hybridization-induced 
Kondo
insulator, where the hybridization gap opens up {\it after} local dynamical
correlations have generated a charge-pseudospin quenched analogue of the 
Kondo
screened state.  The appearance of the high energy Hubbard bands is an 
explicit
manifestation of relevance of local dynamical correlations. At lower 
energies,
hybridization opens up a (renormalized) band insulating gap. Interestingly,
this is a new type of insulating state, which we dub a {\it Charge
Kondo Insulator} (CKI). Obviously, since $G_{ab}(\omega)=\langle f_{a};
f_{b}^{\dag}\rangle$ is finite, this CKI state supports a finite excitonic
polarization in the sense of Portengen {\it et al.}~\cite{sham}.

In case ($ii$), lack of particle-hole symmetry, i.e $\epsilon_{a}'\ne 0$,
effectively leads to partially filled $a,b$ bands, inducing a correlated
metallic state within DMFT. In this case, the $T=0$ spectral function of
the $b$ (``heavy'') fermion shows a sharp quasiparticle resonance
near $\omega=0$, and examination of the self-energy shows that the metallic
state is a renormalized Fermi liquid (FL). "Related signatures of the CKI to
CKL transition are also clearly visible in the ``conduction'' ($a$) band 
DOS."
Thus, DMFT yields a correlated FL metal due to dynamical quenching of the
charge-pseudospin degree of freedom by itinerance of the $b$-fermion (due
to ``hybridization'', $\Delta$, in Eq.~(\ref{eq1})). Emergence of this
charge-Kondo-FL (CKL) within DMFT for the $U_{ab} <0$ PAM is understood
as follows. In the impurity version of the lattice problem, we have an
X-Ray edge problem with recoil (Eq.~(\ref{eq2})) for $\Delta\ne 0$. Within
DMFT, the infra-red singularities described above for $\Delta=0$ are
immediately cut off {\it below} a low energy scale associated with this
recoil. Expressing the low-energy part of $H$ in terms of collective,
{\it bosonic} variables~\cite{mh}, this coherence scale, below which
correlated FL behavior sets in, is estimated to be $k_{B}T_{coh}=Z\simeq$
exp$(U_{ab}^{2}ln(\gamma)/(1-\gamma^{2}))$ with $\gamma=m_{fa}/m_{fb}$,
the mass ratios of the $f_{a,b}$ fermions. In our DMFT calculation, we
find $Z\simeq 0.3$ from the real part of the self-energy (not shown).
Hence, with $t_{b}=0.3t_{a}$, we obtain $T_{coh}\simeq 0.07T_{F}$.  With a
slightly smaller value of $U_{ab}/W$, $T_{coh}$ can be increased to 
$0.1T_{F}$,
which would lie in the lower range of temperatures observable in current
experimental realizations~\cite{esslinger}. Above
$T_{coh}$, the irrelevance of the ``hybridization'' ($\Delta$) implies that
the system crosses over {\it smoothly} to the IBM phase found for 
$\Delta=0$. 
Away from $t_{ab}=\sqrt{t_{aa}t_{bb}}$, the $f_{b}$-fermions once again
acquire a finite mass, giving correlated FL behavior below $T_{coh}$, or
broken symmetry ground states.

{\it Only} if $T_{coh}>T_{SC,CDW}$ will the CKL state be observable in a
small window in parameter space.
%This may be the reason why such a FL-like state has not been observed in
%$Ba_{1-x}(K,Pb)_{x}BiO_{3}$~\cite{AT_96} above 
%$T_{SC}$~\cite{comment,mooij}.
%It appears that $PbTe$ (and $SnTe$) satisfy this condition, and the CKL
%is observed above $T_{SC}$~\cite{gebal}.
The above analysis suggests a way to observe the CKL state experimentally
in optical lattices: 

(a) force a mixture of two atomic states via a Feshbach
resonance (creating attractive interaction), 

(b) apply sequential,
time-synchronized laser excitation to simulate the ``hybridization'', and

(c) ``tune'' $\Delta$ to relevance in the RG sense, by manipulating the
one-body potential~\cite{mandel}. 

The change in the Fermi surface (FS) as
the system is tuned across $t_{ab}=\sqrt{t_{aa}t_{bb}},\Delta=0$ could be
detected by time-of-flight measurements~\cite{fsoptat} for $T>T_{SC,CDW}$.
At $t_{ab},\Delta=0$, a single smeared FS, corresponding to itinerant
$f_{a}$ quasiparticles, should be seen, while, for
$t_{ab}>\sqrt{t_{aa}t_{bb}}$ and/or $\Delta\ne 0$, {\it two} well-defined
FS sheets (${\bf k}$-space eigenvalues of the free part of $H$ in
Eq.~(\ref{eq1}), corresponding to {\it coherent} quasiparticle
propagation, should be observable.  Since the self-energy is strictly {\it
local} in DMFT, the FS will be unaffected by interactions, which then have
the sole effect of critically (or overcritically) damping the FS. Given
that the CKL state involves local, dynamical renormalization, designing
systems where the actual, measured FS agrees well with that computed
without interactions will show improved chances of experimentally
observing the CKL state.

Recently, the quasiparticle dispersion and the spectral function of strongly
interacting fermions exhibiting a BCS-BEC crossover in an optical trap 
has been
experimentally deduced by Stewart {\it et al.}~\cite{jin}. We suggest that
similar experiments carried out below $T_{coh}\simeq 0.07T_{F}$ should 
exhibit
a relatively sharp low-energy peak in $A_{k}(\omega)=
(-1/\pi)$Im$[1/(\omega-\epsilon_{ka}-\Sigma_{a}(\omega))]^{-1}$: this would
constitute a characteristic signature of the CKL state.  The CKI state 
in the
symmetric limit above could also be probed using this technique: one should
observe two ($a$ and $b$ bands) dispersive features below {\it and} above
$E_{F}$, with a finite {\it hybridization} gap at $E_{F}$.  At higher $T$
currently accessible, we predict an incoherent pseudogap structure in
$A_{k}(\omega)$ at low energy. Achieving lower $T$ may be feasible in the
near future: theoretical suggestions to do this by efficient shaping of
the optical lattice profile is predicted to lower the entropy per particle
by a factor of $10$~\cite{kollath}.

Our study also permits drawing an analogy with the phase diagram of
nuclear matter. In quantum chromodynamics (QCD), where {\it three} quarks
form a {\it trion} bound state, a {\it baryon} in the $N=3$ (color) theory,
one also has, as a function of chemical potential, ``color superfluid''
(CSS) and high-$T$, quark matter phases. In the cold-atom setting, in a
pioneering work, such a trion gas phase has theoretically been shown to
be separated from a color superfluid phase by a {\it second order}
transition~\cite{Honerkamp}. Let us now study the possibility of realizing
analogous physics in our case. Our model corresponds to the ``$N=2$''
version of QCD, where the CSS ($p-p$ order) phase is separated from the
excitonic ($p-h$ order) and incoherent IBM phases by a line of
{\it first order} transitions, ending in a second order end-point,
generically at finite $T$. Interestingly, the IBM phase has co-existing
single-particle (``quark'') and pair states at low-energy; it is an
incoherent state where ``quarks'' and ``pairs'' continuously transmute
into each other, their equilibrium fraction being determined by the ratio
$U_{ab}/t, \Delta$ and $T$. The confinement-deconfinement transition, 
familiar
in the QCD context, can be illuminated in our case by employing bosonization
of the ``impurity model'' of DMFT.  At $\Delta=0$, the impurity version of
$H_{FKM}$ is readily {\it bosonized}~\cite{schotte} into a collection of
radial bosonic models about the ``impurity''.  This reads
$L_{0}=v_{F}\int [\Pi^{2}(r)+(\partial_{r}\phi(r))^{2}]dr$, where 
$v_{F}=2ta$
is the carrier fermi velocity. For small
$\delta=(t_{ab}-\sqrt{t_{aa}t_{bb}})\ne 0(<0)$, the $f_{a}-f_{b}$ 
hybridization
is small enough that the {\it two-particle} hopping terms (second order in
$\delta$) are more relevant.  These terms are $H_{res}\simeq
-(\delta^{2}/U_{ab})\sum_{<i,j>}f_{ia}^{\dag}f_{ib}^{\dag}f_{jb}f_{ja}$, 
which
can be decomposed in two ways in a Hartree-Fock decoupling (exact in
$D=\infty$, and in the corresponding impurity model, this corresponds to two, 
$p-h$ and $p-p$ pairing fields), generating either excitonic (particle-hole) order
(in the imbedded impurity problem of DMFT) with order
parameter $\Delta_{ph}=\sum_{i}\langle
f_{ia}^{\dag}f_{ja}+f_{ib}^{\dag}f_{jb}\rangle >0$
or CSS (particle-particle) order, i.e, $\sum_{i}\epsilon^{ab}\langle
f_{ia}^{\dag}f_{ib}^{\dag} \rangle >0$. Bosonization then yields a
{\it dual cosine} quantum sine Gordon (QsG) model:

\be
L= L_{0}-g_{1}\int cos(\beta_{1}\phi(r))dr -g_{2}\int
cos(\beta_{2}\theta(r))dr
\ee
with $\beta_{1}=\sqrt{8\pi K}$ and $\beta_{2}=\sqrt{8\pi/K}. \,\,
K=[1-(U/2\pi v_{F})]^{-1}$  in the weak coupling limit for our negative$-U$
model exactly at the FK point in the impurity model. A renormalization 
group
analysis of $L$ shows two different phases, depending upon the sign of
$\delta g=(g_{1}-g_{2})$~\cite{gogol}:

($i$) when $g_{1}>g_{2}$, the $\phi(r)$ field acquires a finite expectation
value, giving $p-h$ excitonic order. In QCD lore, this translates into 
$\langle
\bar\psi_{\alpha}\psi_{\alpha}\rangle >0$, with $\alpha=a,b$, i.e, to a 
chiral
{\it meson} phase, and,

($ii$) when $g_{2}>g_{1}$, the $\theta(r)$ field acquires a finite 
expectation value, giving CSS order.  This corresponds to 
$\langle\psi_{a}\psi_{b}\rangle >0$,
i.e, to a color-singlet superfluid in QCD.

($iii$) Interestingly, these phases are separated by a critical curve
$g_{1}=g_{2}$ or $K=1$, where the QsG model becomes {\it self-dual}.  This
is exactly soluble, and the correlation functions are characterized by
continuously  varying, $K$-dependent exponents.

The above meson and CSS phases arise from the IBM (bosonic gaussian 
model) via
Kosterlitz-Thouless transitions (driven by relevance of either of the
cosine terms for $K<1$ or $K>1$), and are separated from each other by a
line of first-order transitions, ending in a second-order endpoint at
finite $T=T^{*}$.  For $T>T^{*}$, {\it both} $\langle \bar\psi \psi\rangle,
\langle\psi\psi\rangle = 0$, corresponding to ``quark matter'' in QCD
\cite{hands}. Of course, this analogy is only suggestive: in our case
the $a,b$ particles have very different masses, in contrast with the
small mass difference between $u,d$ quarks in QCD. Nevertheless, the
analogy is interesting, and generalization to $N=3$ is possible.  We
will extend this qualitative analogy in more depth in future.

In conclusion, from a DMFT calculation, we propose that a combination of
wilfully tunable Feshbach resonances {\it and} sequentially applied laser
pulses in optical lattices to tune $U_{ab}/t_{ab}, \Delta$ through a 
desired
range, could facilitate the observation of the elusive CKL state.
Time-of-flight measurements can, by probing the ``Fermi surface'',
provide valuable guidance on the experimental conditions favoring
observation of the CKL state.  Our analysis also shows interesting
analogies with the ``phase diagram'' of nuclear matter.  Extensions to
{\it asymmetric} interaction models~\cite{regal}, allowing for realization
of more exotic ordered phases and {\it anisotropic} CKL states within the
DMFT approach, is also foreseen.

\acknowledgments
M.S.L. and A.T. thank R. M\"ossner and M. Haque for discussions. We 
acknowledge 
helpful comments from T. V. Ramakrishnan and G. V. Pai, and thank C. Kollath
for pointing out Refs.~\cite{jin,kollath}. Financial support and hospitality
from the MPIPKS, Dresden, is gratefully acknowledged.

\end{document}